\documentclass{WileyMSP-template}
\begin{document}

\pagestyle{fancy}
\rhead{\includegraphics[width=2.5cm]{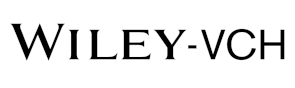}}

\title{Millimeter-wave WISP search with coherent Light-Shining-Through-a-Wall towards the STAX project}

\maketitle


\author{Akira Miyazaki}
\author{Tor Lofnes}
\author{Fritz Caspers}
\author{Paolo Spagnolo}
\author{John Jelonnek}
\author{Tobias Ruess}
\author{Johannes L. Steinmann}
\author{Manfred Thumm}


\dedication{}

\begin{affiliations}
Dr. Akira Miyazaki\\
Msc. Tor Lofnes\\
Department of Physics and Astronomy, Uppsala University, Uppsala, Sweden\\
Email Address: Akira.Miyazaki@physics.uu.se

Dr. Fritz Caspers\\
European Scientific Institute, Archamps, France; and CERN, Geneva, Switzerland

Dr. Paolo Spagnolo\\
INFN Pisa, Pisa, Italy

Prof. Dr. John Jelonnek\\
Dr. Tobias Ruess\\
Dr. Johannes L. Steinmann\\
Prof. Dr. Manfred Thumm\\\
Karlsruhe Institue of Technology, Karlsruhe, Germany 

\end{affiliations}


\keywords{dark photons, millimeter waves, superheterodyne}

\begin{abstract}

A dark photon is one of the simplest extensions of the Standard Model of particle physics and can be a dark matter candidate.
Dark photons kinetically mix with ordinary photons.
The mass range from $10^{-4}$ to $10^{-3}$~eV of such dark photons is under-constrained by laboratory-based experiments and a new search is therefore motivated.
In this mass range, dark photons behave like waves rather than particles and the corresponding electromagnetic waves are in the millimeter-wave range.
The technical difficulties of the millimeter waves have prevented so far dark photon experiments in this mass range.
We propose the use of coherent millimeter waves to search for dark photons in a Light-Shining-through-a-Wall (LSW) experiment.
We clarify the merit and limitations of coherent wave detection and briefly investigate the potential of single photon sensors at microwaves.
Development of millimeter-wave technology is not only limited to dark photons.
Technically, an experiment for dark photons by using electromagnetic waves resembles that for axions, another light dark matter candidate, with static magnetic fields.
This paper represents an essential step towards axion LSW in the millimeter-wave range (STAX experiment) as a potential successor of an on-going experiment in infrared.
\end{abstract}


\section{Introduction}
The discovery of the Higgs boson in 2012~\cite{201230, 20121} was a major milestone and a confirmation of the Standard Model (SM) of particle physics. 
Despite its success, the investigation of extensions to the SM continues to be a highly compelling area of research within the field of particle physics. 
One possibility is the existence of a {\it dark sector} containing of new particles that do not have gauge interactions with the known particles in the SM~\cite{PhysRevD.79.015014, Nima-Arkani-Hamed_2008, PhysRevLett.101.231301}. 
These new particles could include scalars, pseudoscalars, and sterile neutrinos. 
One example of a new particle is the so-called {\it dark photon}~\cite{HOLDOM1986196}, a hypothetical particle that would couple to the SM photon via kinetic mixing.
However, there is currently no experimental evidence for these new particles.

Astronomical observations~\cite{1970ApJ...159..379R, Markevitch_2004, Clowe_2004} suggest that the Universe is filled with dark matter interacting with ordinary matter only through gravity.
Beyond-the-Standard-Model particles could explain the dark matter puzzle.
Amang these particles, there are dark photons.
This work aims to search for dark photons in a mass range from $10^{-4}$ to $10^{-3}$~eV with a Light-Shining-through-a-Wall (LSW) experiment that is free from systematic uncertainties in astronomical observations.
This mass range has not been well constrained~\cite{6380414, 9370386} by previous laboratory-based experiments due to technical difficulties in millimeter waves.
We use coherent wave detection method, as opposed to photon counting detectors, as these have a better energy resolution within this mass range.
While this technique is not novel, its application in particle physics experiments has been limited.
The LSW experiment enables strong temporal coherency in dark photon unlike astronomical dark photons.

In LSW experiments, the advantage of the coherent wave detection was recognized rather recently in the infrared laser experiment (ALPSII)~\cite{Isleif2022, PhysRevD.99.022001,https://doi.org/10.48550/arxiv.2010.02334}.
In this work, we show the feasibility of the coherent detection technique for the search of dark photons in the millimeter-wave range.
Although a similar concept was verified at 1-3~GHz~\cite{Caspers:1195741, F_Caspers_2009, PhysRevD.88.075014}, an experiment at 30~GHz is novel.
Note that microwave technology is not just naively scaled in this frequency range.

In the next section, we develop a theory of LSW for dark photons at 30~GHz.
We describe the quantum coherent states and their implication in particle counting and classical wave detection.
After that, we show the results of proof-of-concept experiments.
Future prospects include the discovery potential of this method and revisit the fundamental limitation of coherent wave detection.
We briefly mention the larger scale experiment for axions in the future (STAX).
The last section is dedicated to concluding remarks.

\section{Theory of coherent dark photon detection}
\subsection{Classical theory of dark photons and the LSW experiment}
We consider dark photons which kinetically mix with ordinary photons with a small mixing parameter $\chi$.
We perturbatively take into account the mixing at the first order~\cite{JAECKEL2008509}.
The vector potential ${\bf A}(t, {\bf x})$ of ordinary photons generates the dark vector potential ${\bf A'}(t, {\bf x})$ in a modified Proca equation (vector version of Klein-Gordon equation) which satisfies
\begin{equation}
\left( \frac{\partial^2}{\partial t^2} - {\bf \nabla}^2 + m_{\gamma'}^2 \right) {\bf A'}(t, {\bf x}) = \chi m_{\gamma'}^2 {\bf A}(t, {\bf x}), \label{eq:photon_to_dp}
\end{equation}
with $m_{\gamma'}$ being the dark photon mass.
In a same way, ${\bf A'}(t, {\bf x})$ is a source term of a modified Maxwell equation
\begin{equation}
\left( \frac{\partial^2}{\partial t^2} - {\bf \nabla}^2 +\chi^2m_{\gamma'}^2 \right) {\bf A}(t, {\bf x}) = \chi m_{\gamma'}^2 {\bf A'}(t, {\bf x}), \label{eq:dp_to_photon}
\end{equation}
which indicates that the kinetic mixing of dark photons provides a very small effective mass $\chi^2m_{\gamma'}$ to ordinary photons.
Therefore, dark photon mixing can also be studied by precision tests of Coulomb's law~\cite{PhysRevLett.61.2285}.

Equations~(\ref{eq:photon_to_dp}) and (\ref{eq:dp_to_photon}) lead to photon-dark-photon oscillations because their mass eigenstates are rotated from that of U(1) gauges.
Quantum oscillation of this kind can be tested by preparing one of the states and observing the same state at some distance, as is done in neutral meson oscillation~\cite{1955PhRv...97.1387G} and neutrino oscillation experiments~\cite{1968JETP...26..984P}.
In case of dark photon physics, one needs to prepare very intense photon fields confined in a bounded space, such as a resonant cavity,
so that due to photon-dark-photon oscillation, some of the initial photons may be converted to the dark photons and leak through the boundary of the photon field.
Reconverted photons may be detected by a very sensitive photon sensor or wave detector.
This is the working principle of LSW experiments.

The LSW technique has pros and cons compared to direct searches for dark matter dark photons.
An advantage is that one can be free from astrophysical uncertainties in the dark matter density and model dependence in their coherency.~
In principle, LSW depends only on an equation of motion of dark photons (Eq.~(\ref{eq:photon_to_dp})).
One can fix strength of the source field and experimentally define the coherency of dark photons.
However, in LSW experiments, the conversion happens twice, i.e. a photon to a dark photon and a dark photon to a photon, so that the sensitivity of the search is reduced by $\chi^4$~\footnote{The mixing parameter $\chi$ is defined with field amplitudes of dark photons in the Lagrangian. The measurement sensitivity depends on intensity that is proportional to $\chi^2$. Since LSW has twice the conversion, the over-all sensitivity is proportional to $\chi^4$.}.
Also, one needs to handle purely electromagnetic leakage of photons (cross-talk) without dark photon conversion.
Therefore, small signal detection and background reduction are more challenging than dark matter searches.

The concepts of a photon as a particle and as an electromagnetic wave have been conflated thus far.
Consequently, the question arises as to how the classical wave equations, as represented by Eq.~(\ref{eq:photon_to_dp}) and (\ref{eq:dp_to_photon}), can be interpreted within the framework of a quantized field.
To establish the validity of classical formalism, it is necessary to introduce the coherent state, a well-established concept in quantum optics.
This will enable us to better understand the relationship between classical wave equations and quantized fields and further clarify the nature of photons as both particles and waves.

\subsection{Coherent state and classical waves}
A key in quantum optics is the coherent state originally introduced by Schr\"{o}dinger~\cite{Schrodinger1926}.
In this section, we review the formalism established by Glauber~\cite{PhysRev.131.2766}.
An electric field operator with an angular frequency $\omega$ in quantum optics can be written as
\begin{equation}
\hat{E} (t, {\bf x})  = \hat{E}^{+} e^{-i(\omega t-{\bf k}\cdot{\bf x})} +  \hat{E}^{-} e^{+i(\omega t-{\bf k}\cdot{\bf x})}, \label{eq:field_operator}
\end{equation}
where the two operators $\hat{E}^{+}$ and $\hat{E}^{-}$ are defined as operators for $\exp(-i\omega t)$ and $\exp(+i\omega t)$ waves, respectively.
A coherent state is defined as an eingenstate of an annihilation operator
\begin{equation}
\hat{a}|\alpha\rangle = \alpha | \alpha \rangle,
\end{equation}
where $\alpha$ is a complex number whose absolute value squared $\left|\alpha \right|^2$ represents the mean number of photons.
The link to the classical wave is revealed by the fact that an electric field operator for $\exp(-i\omega t)$ is proportional to $\hat{a}$
\begin{equation}
\hat{E}^{+} = i\sqrt{\frac{h\nu}{2\epsilon_0 V}} \hat{a}
\end{equation}
with $V$ the quantization volume and thereby the expectation value of $\hat{E} (t, {\bf x})$ is
\begin{equation}
\langle \alpha | \hat{E} (t, {\bf x}) | \alpha \rangle = E^{+} e^{-i(\omega t-{\bf k}\cdot{\bf x})} +  E^{-} e^{+i(\omega t-{\bf k}\cdot{\bf x})}, \label{eq:coherent_expectation}
\end{equation}
where we introduced complex values $E^{+}$ and $E^{-}$.
This Eq.~(\ref{eq:coherent_expectation}) is the solution of the classical Maxwell equation.
To the linear order, the conclusion derived from Eq.~(\ref{eq:photon_to_dp}) and (\ref{eq:dp_to_photon}) is applied to the expectation value of these field amplitudes.
In Appendix~\ref{sec:photon_counting}, we summarize how the photon counting works on the coherent states.

\subsection{Phase-lock and relative temporal coherency}
\begin{figure} [h]
\begin{center}
  \includegraphics[width=0.8\linewidth]{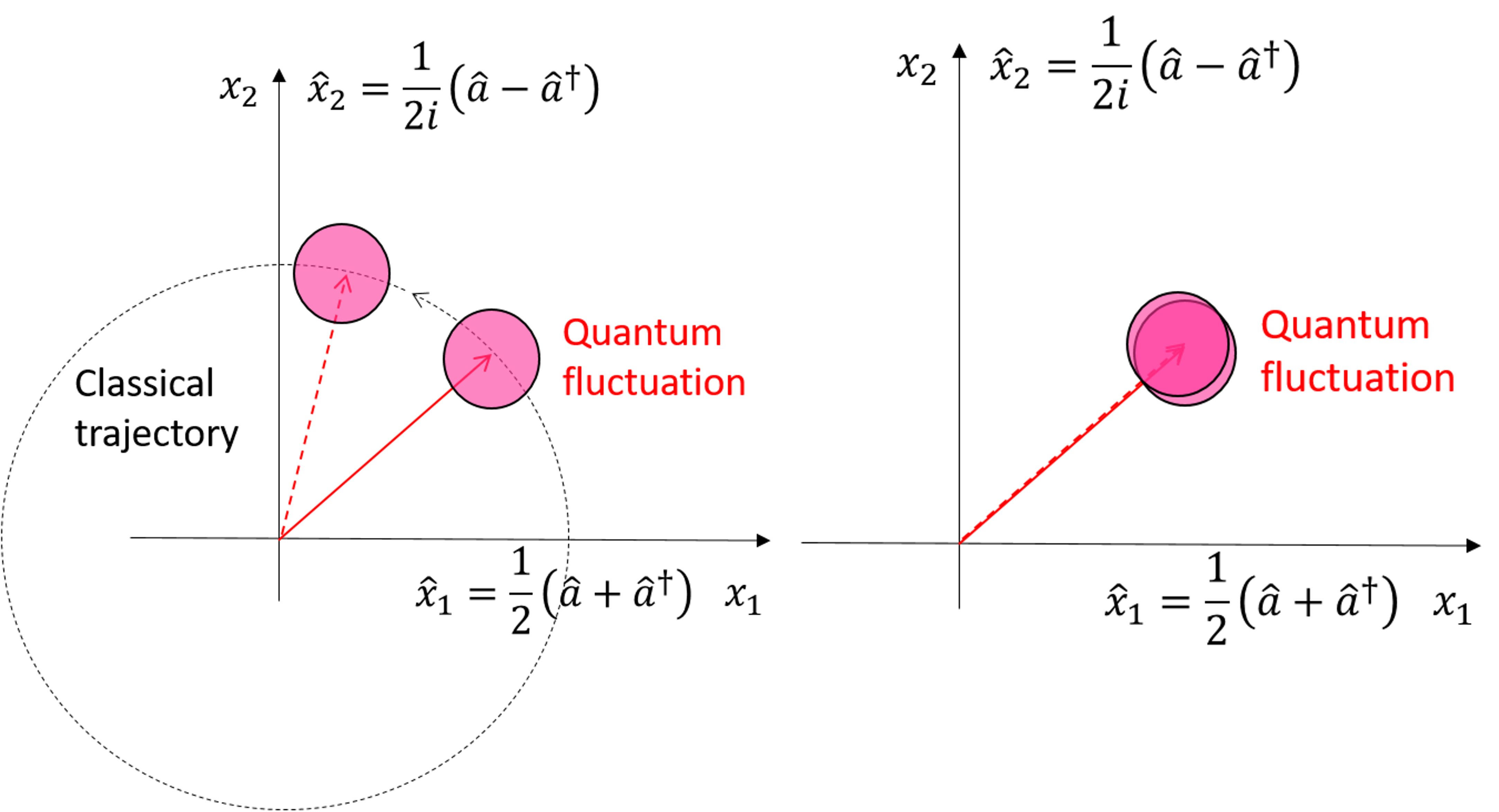}
  \caption{Schematic of phase locking of coherent states: without lock (left) and with lock (right). The horizontal axis shows an amplitude of one of the phase components and the vertical axis shows the other. The pink colored small circles illustrate the coherent states projected onto these amplitudes with the quantum uncertainty. The center of the small circle, the expectation value of the coherent state, rotates on the larger dashed circle, representing the classical phase shift due to thermal drift of the devices. This classical drift is suppressed in the right figure thanks to the phase locking.}
  \label{fig:lock}
\end{center}
\end{figure}

A coherent state is a very good model of photon fields from lasers in the range of optical frequencies and a chain of a signal generator and amplifiers in a microwave circuit.
An ideal {\it relative} temporal coherency can be achieved with the phase-locking method.
Figure~\ref{fig:lock} shows a schematic of the phase locking mechanism.
The quantum states of a generator and a detector are visualized on axes that represent two independent classical amplitudes of $\sin(\omega t)$ and $\cos(\omega t)$ components.
In free running operation of these two devices, an expectation value of each photon field, the center of the small circles, independently rotates along the circle with a constant radius $\sqrt{(E^{+})^2 + (E^{-})^2}$.
This phase shift corresponds to thermal drift of internal reference quartz oscillators in each microwave device; therefore, it is a classical phenomenon.
Here, the fundamental quantum fluctuation is shown as a small circle around the expectation value.
In Appendix~\ref{sec:wave_detection}, we summarize how the wave detection works on the coherent states.
We disregard potential squeezing of the quantum states in this paper for simplicity.

By introducing a phase-lock between the generator and the detector, the relative phase shift between them can be strongly reduced.
Consequently, the expected signal becomes artificially a $\delta$-function.
Since the major background events are white noise from black body radiation and Johnson-Nyquist noise, 
the signal-to-noise ratio can be dramatically improved.
This phase-locking can be implemented naturally by the standard 10~MHz reference line in a microwave measurement equipment.

Importantly, the quantum field of dark photons is assumed to almost perfectly preserve the same coherency as the photon fields~\cite{9567608}.
This is implied by the coupled {\it classical} wave equations Eq.~(\ref{eq:photon_to_dp}) and (\ref{eq:dp_to_photon}).
Quantum mechanically, this description simply influences only the expectation values of field operators.
The propagation of dark photons with uncertain mass introduces a small decoherence and the frequency distribution is smeared from the $\delta$-function~\cite{F_Caspers_2009}.
It is possible that the electric noise within the phase-locking circuit at high frequencies of up to 30~GHz determines the narrowest bandwidth.
Consequently, prior to developing a setup for the search of dark photons, experimental validation is crucial.

\section{Validation of coherent wave detection scheme at 30~GHz}
The relative temporary coherency was validated in the CERN Resonant WISP Search (CROWS) experiment~\cite{PhysRevD.88.075014} at lower frequency (1.7~GHz in the TM$_{010}$ mode and near 3~GHz in the TE$_{011}$ mode).
Two of the authors (F.~Caspers and M.~Thumm) were involved in the CROWS project.
In order to address the under-constrained mass region ($10^{-4}$-$10^{-3}$~eV), this work introduces two novel ideas beyond those developed in the CROWS experiment:
\begin{enumerate}
\item Coherent millimeter wave system at higher frequency 30~GHz 
\item Quasi-optical oversize-resonators at 30~GHz instead of conventional resonant cavities near the fundamental mode
\end{enumerate}
This paper reports on point 1 of this project about the coherency.
Depending on the implementation of up-conversion in the experimental arrangement, 
there may be some technical limitations in the temporal coherence of 30~GHz devices.
Therefore, experimental tests at 30~GHz are essential as a proof-of-principle.
The resonator design in point 2 will be presented elsewhere.

\subsection{Relative temporal coherency of the signal}
\begin{figure} [h]
\begin{center}
  \includegraphics[width=0.9\linewidth]{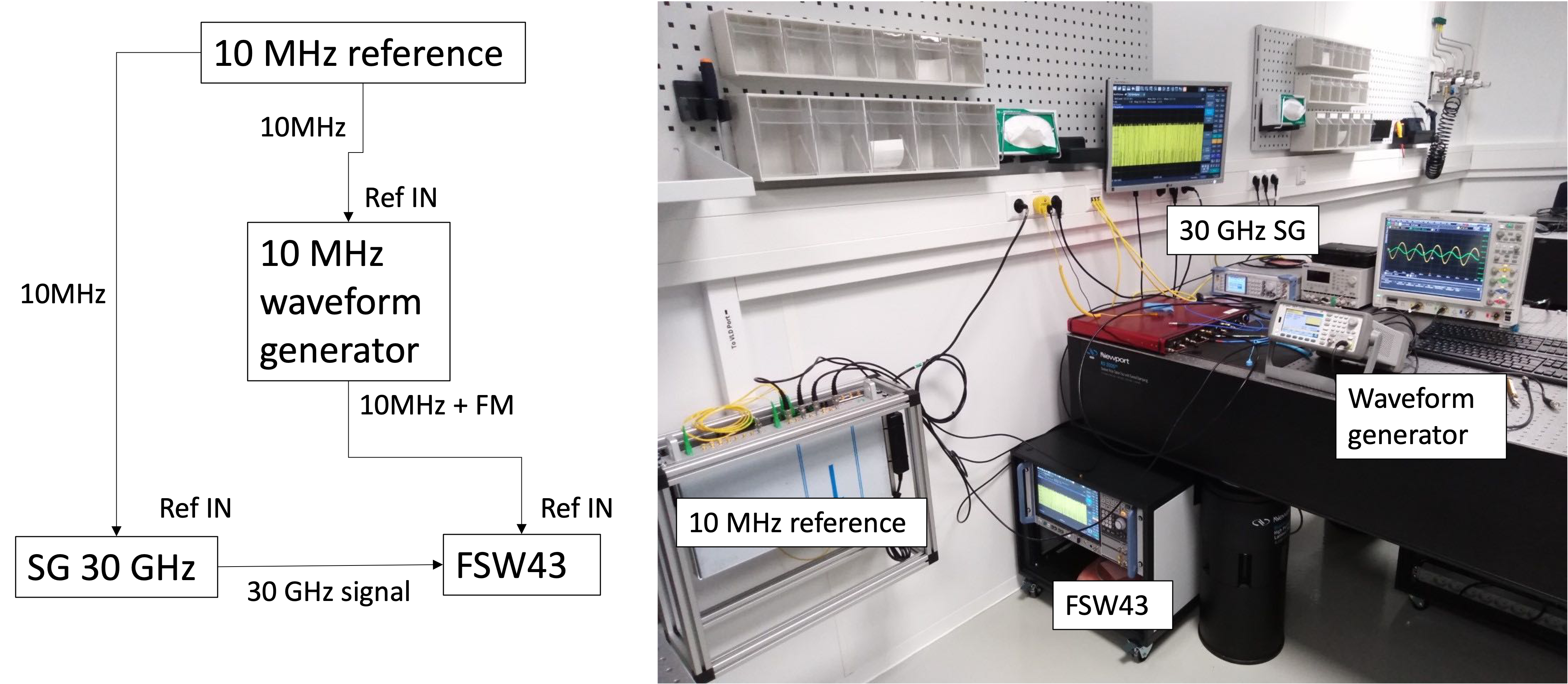}
  \caption{Test setup of validation of the relative coherency. The left figure shows the circuit diagram and the right shows the photograph.}
  \label{fig:setup_coherency}
\end{center}
\end{figure}

We performed a simple experiment to test the idea of relative coherency at 30~GHz~\cite{Miyazaki:2798112}.
We used a real-time spectrum analyzer (FSW43; R\&S)~\cite{9896105} and a digital signal generator (SMB100A; R\&S) located in the THz laboratory at the Karlsruhe Research Accelerator (KARA)~\cite{KARA}.
Figure~\ref{fig:setup_coherency} shows the experimental setup.
The high-quality 10~MHz reference is provided by the facility.
The signal generator is phase-locked to this reference.
The 30~GHz output from the signal generator is fed into the analyzer.
The analyzer is phase-locked to a 10~MHz signal through a waveform generator (Keysight; 33521A) to add artificial phase noise or offset to the reference.
The waveform generator was connected to the reference line of the signal generator instead when the generator's response was tested.

The first test was to check the response of the 30~GHz signal against an offset in the 10~MHz reference line.
This test reveals the method of up-conversion in the apparatus.
When an offset of $\pm1$~Hz was added to the 10~MHz reference line of the analyzer,
the monitored 30~GHz signal frequency changes by $\mp3$~kHz.
These results indicate that a relative change in the reference line is linearly propagated to the signal
\begin{equation}
\frac{1 {\rm \, Hz}}{10 \, {\rm MHz}} = \frac{3\, {\rm kHz}}{30\,  {\rm GHz}}.
\end{equation}
When an offset of $\pm1$~Hz was added to the generator's reference line instead,
the monitored 30~GHz signal frequency changes by $\pm3$~kHz.
Since a frequency shift in the generator has opposite polarity to that in the analyzer,
importantly, a natural drift, vibration, or jitter in the 10~MHz reference line is cancelled when the same reference line is shared by both devices.
However, if a jitter in the common reference line becomes larger than the phase-lock-loop circuit can handle, 
the relative coherency is lost during the response time of the circuit.
We tested this by adding 1~Hz frequency modulation in the 10~MHz common reference line to both devices.
The phase-locking, and therefore the relative coherency, of these devices was recovered within 10~ms.
In reality, a commercially available 10~MHz RF signal has much better stability.
The intrinsic decoherence from the jitter in the reference line gave better side-bands around the 30~GHz signal than an artificially added frequency modulation of 1~$\mu$Hz on 10~MHz.

\begin{figure} [h]
\begin{center}
  \includegraphics[width=0.6\linewidth]{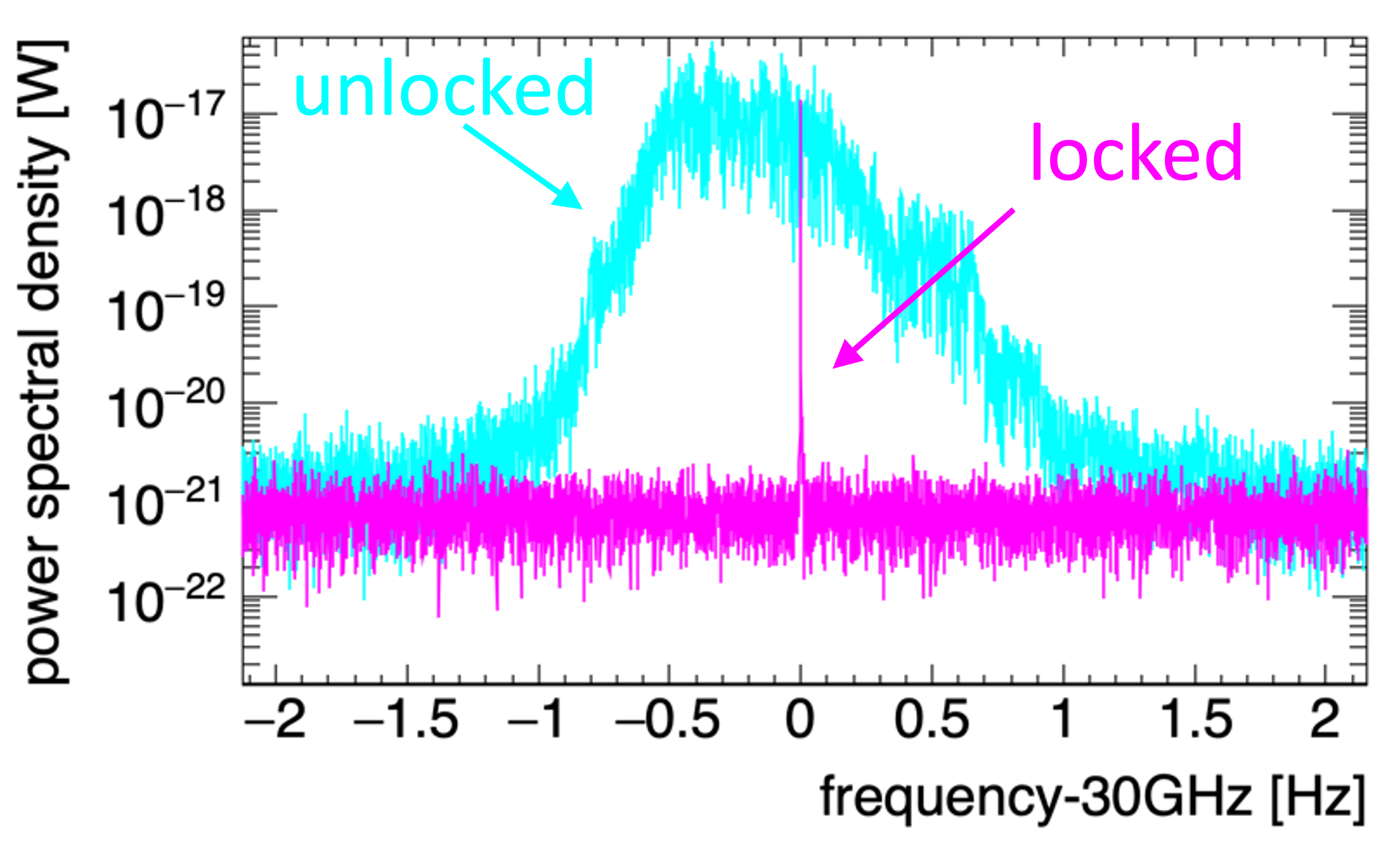}
  \caption{Demonstration of phase locking. The blue plot shows free running spectrum without the phase lock. The magenta plot shows virtually a $\delta$-function thanks to the phase-lock scheme investigated in this work.}
  \label{fig:lock_spectrum}
\end{center}
\end{figure}

Figure~\ref{fig:lock_spectrum} compares the 30~GHz signal integrated over 5240~s with and without the common 10~MHz reference.
Note that an offset of 55~Hz out of 30~GHz on the unlocked case is manually corrected in the plot.
The data clearly show that the classical drift of the individual devices is corrected and $\delta$-function-like relative coherence is achieved, just as illustrated by Fig.~\ref{fig:lock}.
After integrating the data over 5240~s, we achieved the resolution bandwidth of 169~$\mu$Hz and the signal bandwidth was within one bin of it with a marginal side-band three orders of magnitude lower than the center signal at 30~GHz.
This argument is valid up to the integration time of $t<t_{\rm co}=1/\Delta\nu_{\rm \gamma'}$ the relative coherent time of the setup.
In this work, we did not observe a practical limit of $t_{\rm co}$ up to 5240~s.

\subsection{Narrow-band filtering of thermal noise} \label{sec:narrow_band}
The benefit of coherency is the ultimate energy resolution and therefore linear enhancement of the signal-to-noise ratio thanks to narrow-band noise filtering.
The real-time spectrum analyzer in the setup Fig.~\ref{fig:setup_coherency} has a typical noise temperature of $T_{\rm s}=2.3\times10^5$~K, which corresponds to a power spectral density of
\begin{equation}
k_BT_{\rm s} = -145\, {\rm dBm/Hz} = 3.2\times 10^{-18}\, {\rm W/Hz}.
\end{equation}
By integrating the noise over longer than 1~s, one can narrow down the resolution bandwidth $\Delta\nu_{\rm RBW}$ below 1~Hz.
Thus, one could further filter out the noise power from this intrinsic noise level of the device in the special case of the LSW setup.

From Eq.~(\ref{eq:noise_power_linear}), the noise power of this frequency range ($h\nu\ll k_{\rm B}T$) is well described by Dicke's radiometer formula
\begin{equation}\label{eq:Dicke}
P_{\rm N} = k_{\rm B} T_{\rm s} \sqrt{\frac{\Delta\nu}{t}}.
\end{equation}
where $\Delta\nu$ is the bandwidth relevant to the measurement configuration and $t$ is the integration time.
For example, if the signal contains  incoherent photons, $\Delta\nu$ is determined by the energy resolution of the calorimeter or an antenna or an analog filter.
This noise simply reflects the standard deviation of the Poisson distribution from photon counting.
If the signal contains coherent waves, $\Delta\nu$ is determined by either the signal bandwidth $\Delta\nu_{\rm \gamma'}$ or resolution bandwidth $\Delta\nu_{\rm RBW}$ of the spectrum analyzer.
Since a phase-locking technique providing a signal bandwidth narrower than the resolution bandwidth i.e. $\Delta\nu_{\rm \gamma'}<\Delta\nu_{\rm RBW}$,
\begin{equation}
\Delta\nu = \Delta\nu_{\rm RBW} \equiv 1/t \label{eq:nu_is_over_t}
\end{equation}
is applied in Eq.~(\ref{eq:Dicke}), and the noise power is filtered out linearly by the integration time
\begin{equation}
P_{\rm N} =  \frac{k_{\rm B} T_{\rm s}}{t}. \label{eq:Dicke2}
\end{equation}
Such a long time integration can be implemented by a Fast Fourier Transform (FFT) of waveform sampled over more than 1 second.
This was performed in the real-time spectrum analyzer FSW43.

\begin{figure} [h]
\begin{center}
  \includegraphics[width=0.8\linewidth]{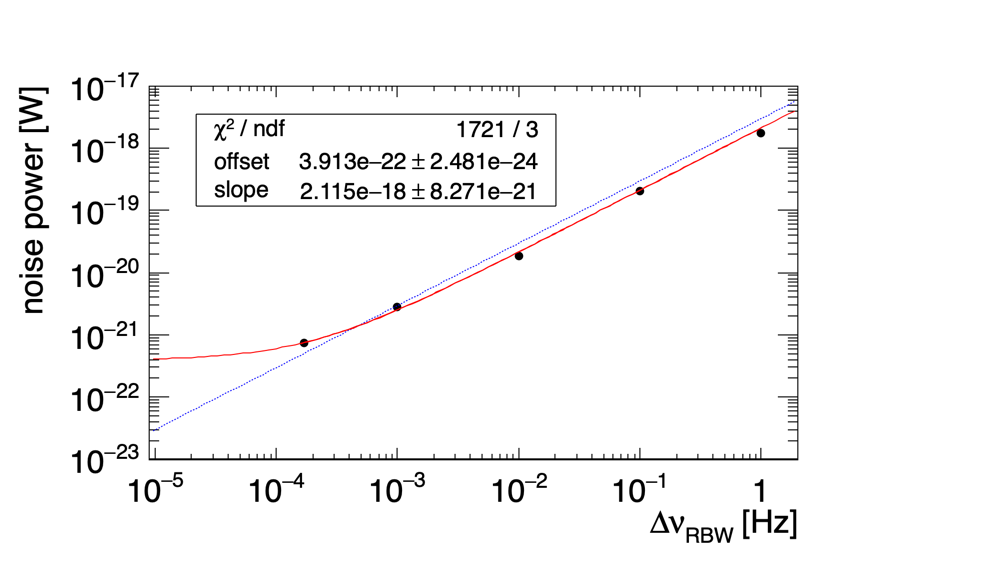}
  \caption{Linear filtering of noise power by resolution bandwidth (RBW). The black dots show the experimental data obtained by fitting power spectral density of side-bands sufficiently away from the signal at 30~GHz. The blue dashed line shows a linear fitting. The red solid line shows linear + offset fitting and it reveals the saturation of noise filtering below 100~$\mu$Hz.}
  \label{fig:PN_vs_RBW}
\end{center}
\end{figure}

Figure~\ref{fig:PN_vs_RBW} shows the noise power, which is an average of the white noise baseline in Fig.~\ref{fig:lock_spectrum} as a function of $\Delta\nu_{\rm RBW}$.
This result demonstrated the successful narrow-band filtering at 30~GHz and also revealed the limitation of the present setup.
The noise power followed the prediction of Eq.~(\ref{eq:Dicke2}) and started to deviate below 100~$\mu$Hz.
We found that the noise filtering was saturated at $3.9\times 10^{-22}$~W.
In conclusion, the narrow-band filtering scheme of the present setup reduced the noise power from $3.2\times 10^{-18}$~W to $3.9\times 10^{-22}$~W.
Importantly, this noise power, dominated by the spectrum analyzer, can be further reduced by a low-noise amplifier chain.

\section{Future prospect}
\subsection{Setup of the lower-power experiment at 300~K}\label{sec:PoP_setup}
\begin{figure} [h]
\begin{center}
  \includegraphics[width=1.0\linewidth]{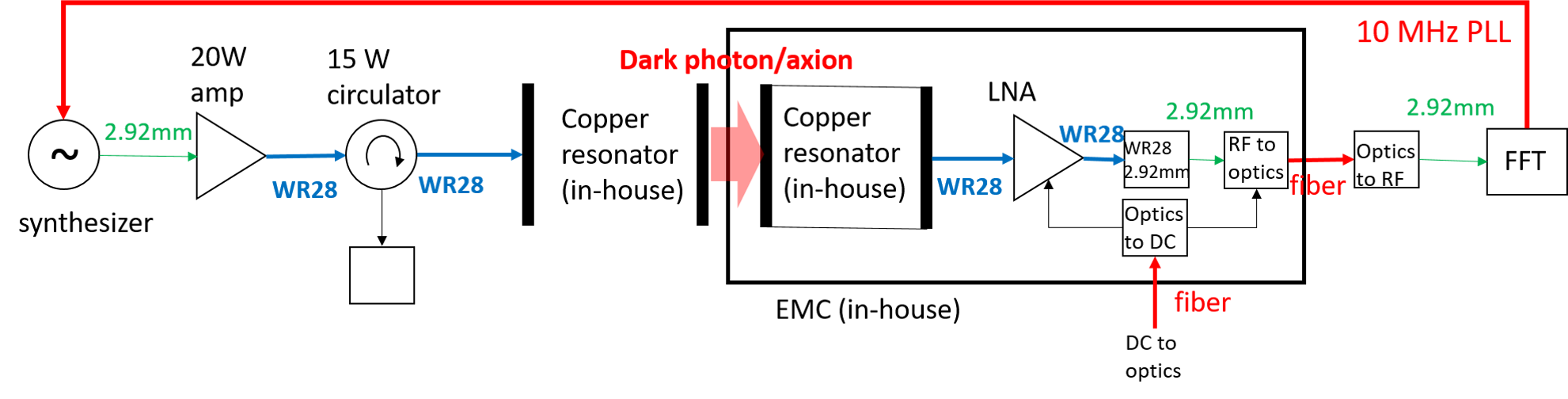}
  \caption{Schematic diagram of proof-of-principle experiment.}
  \label{fig:PoP_schematic}
\end{center}
\end{figure}

Based on the results presented in this paper, we plan to perform a low-power experiment with a commercially available apparatus.
Figure~\ref{fig:PoP_schematic} shows the schematic of this experiment.
The key for this arrangement is a signal generator and a FFT analyzer locked by a common reference signal.
In order to increase the intensity of the millimeter waves, a 20~W amplifier and a Fabry-Perot resonator are used for the dark photon generation.
This amplifier introduces side-bands but does not distort the relative coherency of the signal frequency.
With two mirrors made of pure copper, the finesse of the resonator at 300~K can be around 3300 which intensifies the 30~GHz wave by around $\beta\sim1000$.
This setup can provide effective power of
\begin{equation}
20\, {\rm W} \times 1000 = 2\times10^{4}\, {\rm W}.
\end{equation}
In terms of the mean number of photons, it satisfies
\begin{equation}
\frac{20\, {\rm W} \times 1000}{h\times 30\, {\rm GHz}}=10^{27}
\end{equation}
photons per second. 
Only a very small fraction of such a large number of photons may become dark photons and go to the detection side of the experiment.
Another Fabry-Perot resonator is located at the mirror position at the detection side of this experiment.
The dark photons coming to this resonator are converted back to the original coherent photons, or more explicitly, classical electromagnetic waves as stressed in this paper.
This resonator enhances the signal by another factor of 1000; thus, the expected signal power satisfies
\begin{equation}
P_{\rm s} = 2\times10^{7}\times \chi^4 \, {\rm W}.
\end{equation}
where phase factors as a function of resonator length and wavelength are disregarded for simplicity.

It is crucial to add a low-noise amplifier (LNA) just after this resonator.
A commercially available LNA at 300~K can amplify the signal by 1000 (30~dB) while adding only a system temperature of 110~K.
With this gain of 30~dB, the intrinsic noise level of the FFT analyzer, which was $3\times10^{-18}$ W/Hz in FSW43, can be effectively suppressed down to $10^{-21}$~W/Hz that is comparable to the 300~K thermal noise ($-174$ dBm/Hz) from the resonator and waveguide chain.
Since the system temperature is dominated by the LNA's 110~K, one can even monitor the cavity resonance from 300~K blackbody radiation~\cite{PhysRevD.88.075014}.
Since the narrow-band filtering scheme demonstrated in Sec.~\ref{sec:narrow_band} can reduce the noise power by four orders of magnitude, 
the noise power will be
\begin{equation}
P_{\rm N} = 10^{-25}\, {\rm W}.
\end{equation}
Therefore, the signal-to-noise ratio can be given by
\begin{equation}
S/N = \frac{P_{\rm s}}{P_{\rm N}} = \frac{2\times10^{7}\times \chi^4}{10^{-25}} = 2\times 10^{-32} \times \chi^4.
\end{equation}
In this low-power measurement, one can address the $\chi\sim10^{-8}$ region around $m_{\gamma'}\sim10^{-4}$~eV. 

The above argument is justified if cross-talks of 30~GHz are sufficiently suppressed.
The resonator and the LNA chain must be installed inside a electromagnetic shield in order to avoid cross-talk from the generator side.
A nearly perfect electromagnetic shield is realized by avoiding the use of any coaxial feedthroughs because such feedthroughs are transparent to broadband RF.
Instead, DC electric power for the LNA and the RF signal can be transported by an optical fiber system.
A fiber requires a small hole, not coaxial, and such a hole can be at the signal frequency below cut-off.
With this technique, a shielding efficiency of nearly $-300$~dB was achieved in the CROWS experiment~\cite{PhysRevD.88.075014} at 1-3~GHz.
Since the cut-off diameter is just scaled down by the frequency, a similar shielding for 30~GHz is also feasible.

\subsection{Potential of a high-power experiment}\label{sec:PLL_gyrotron}
The phase-locking between a photon generator, i.e. oscillator, and a wave detector, is key for this project.
In the low-power experiment, phase-locking is straightforward with the standard 10~MHz line connected to a signal generator at 30~GHz.
The linear amplifier introduces phase noise and adds non-linear distortions in the signal that appear as side-bands but is not significant to our experiment.
However, in order to increase the search sensitivity to higher power, we face technical challenges.
Solid-state amplifiers above 30~GHz are commercially available up to 200~W today.
This provides an improvement of a factor of 10 in the number of photons from the low-power setup with a 20~W amplifier described in Sec.~\ref{sec:PoP_setup}.
Since the total conversion efficiency from a photon to a dark photon and back to a photon is proportional to $\chi^4$,
the sensitivity of the search is improved by only $^4\sqrt{10}=1.88$.

In order to reach even higher powers than 1~kW, vacuum electron tubes are an option.
There are two categories in vacuum electronic devices at this frequency range:
\begin{enumerate}
\item amplifiers: klystrons, gyro-klystrons, gyro-TWT
\item oscillators: backward wave oscillator (BWO), gyrotrons, gyro-BWO
\end{enumerate}
One may consider the amplifiers as the first option because the experiment requires excellent relative coherency.
Amplifiers in principle preserve the temporal coherency of the generator, which is phase-locked to the detector, while they add side-bands.
High-power amplifiers around 30~GHz have been developed for the application of accelerators (Ka-band CLIC)~\cite{https://doi.org/10.1049/iet-map.2018.0103, doi:10.1063/1.5144590}.
However, regardless of the very high peak power, such amplifiers are in short pulsed operation and the average power is low.
The search sensitivity of dark photons is determined by the average power.

The oscillators listed above have been developed for Continuous Wave (CW) applications.
In particular, gyrotrons can provide the highest average power in this frequency range.
A gyrotron is a coherent oscillator of millimeter waves based on the cyclotron maser resonance.
Since the cyclotron maser resonance is stable in CW operation,
most typical gyrotrons are in CW operation or long pulsed operation.
The state-of-the-art gyrotrons~\cite{Thumm2020} can generate 1~MW CW power at up to 170~GHz for half an hour and such 24 gyrotrons' outputs are combined to heat up the plasma in the future fusion reactor ITER.
In KIT, an R\&D gyrotron is available at 28~GHz and can generate 20~kW CW for more than 1 hour~\cite{Malygin2014}.
In terms of the number of photons for dark photon searches, gyrotrons are the most promising option.

An issue for gyrotrons has been their somewhat limited temporal coherency.
In a typical free-running gyrotron at 200~GHz, the signal bandwidth has been around 1~MHz ($Q_{\rm \gamma'}\sim2\times 10^5$)~\cite{miyazaki_PhD}.
This gives an amazingly good coherency at this frequency range.
However, this excellent coherency is still not sufficient for the coherent dark photon search.

In order to overcome the limitation in coherency, phase-locked gyrotrons have been developed.
In a gyrotron, the electron beam for the cyclotron maser resonance is emitted from a magnetron injection gun.
This electron gun is either a diode or triode to properly accelerate and guide electrons to the center of a gyrotron cavity.
By adding phase-lock control in the high-voltage power supply of either the cathode or the modulation anode,
the output frequency can be dramatically stabilized.
A recent study achieved 1~Hz {\it absolute} signal bandwidth for a 170~GHz and 25~kW gyrotron~\cite{Denisov2022}.
Locking the phase-lock circuit to the common reference of the FFT analyzer is a promising way to achieve excellent {\it relative} temporal coherence for the dark photon search.


\subsection{Potential of single photon counting}\label{sec:JES}
Cooling down the resonator to 4~K and installing a HEMT amplifier (noise temperature 7~K) can further enhance the signal-to-noise ratio.
However, the coherent wave detection technique is eventually blocked by the standard quantum limit.
In case of 30~GHz electromagnetic waves, one has
\begin{equation}
T_{\rm SQL} = \frac{h \times 30\, {\rm GHz}}{k_{\rm B}} = 1.4\, {\rm K}
\end{equation}
and practically, the gain by cooling down stops around this temperature even if one employs a parametric amplifier to reach $T_{\rm SQL}$.
Moreover, this work revealed a practical limitation of narrow-band filtering down to $\Delta\nu_{\rm RBW}\sim100$~$\mu$Hz in the present setup.

Eventually, photon counting sensors may come into the game even in the LSW experiments.
Appendix.~\ref{sec:dark_photon_coherency} summarises previous arguments in dark matter search.
From Eq.~(\ref{eq:ratio_noise_power}), for any $Q_{\rm \gamma'}\propto\Delta\nu^{-1}$, noise power of single photon sensors surpasses the linear amplifier's noise if everything including a detection cavity is cooled down below $50$~mK.
Although linear amplifiers cannot improve their noise power below $k_{\rm B}T_{\rm SQL}$ even if their physical temperature goes down,
the noise level of the photon sensor is exponentially improved by cooling down because blackbody radiation is dramatically suppressed.
This picture is only valid if the dark counts of the photon sensor is dominated by blackbody radiation.
Development of a single photon calorimeter around 30~GHz is a technically challenging task.
In the National Enterprise for nanoScience and nanoTechnology (NEST) in Pisa,
a Josephson Escape Sensor has been developed~\cite{PhysRevApplied.14.034055}.
This sensor is based on a phase state in a nano-wire Josephson junction under bias current and behaves as a very sensitive transition edge sensor~\cite{doi:10.1063/5.0021996} in case of absorbing a microwave photon.
The noise equivalent power is expected to be $10^{-25}$ W/$\sqrt{\rm Hz}$ at 20~mK.
The next step is to demonstrate this noise level by injecting a very weak microwave signal into a sensor placed inside a dilution refrigerator.
Successful implementation would lead to beyond the state-of-the-art detection of microwave photons in this frequency range.

\subsection{Predicted exclusion limits of dark photons}
Under the plane-wave approximation, the probability of LSW via photon-dark-photon oscillation is
\begin{equation}\label{eq:photon-dark-photon-oscillation}
p(L, \omega) = \left( \frac{\omega + \sqrt{\omega^2 - m_{\gamma'}^2}}{\sqrt{\omega^2 - m_{\gamma'}^2}}\right)^4 \chi^4 \sin^4{\left[\frac{L}{2}\left( \omega - \sqrt{\omega^2 - m_{\gamma'}^2}\right) \right]},
\end{equation}
with $L$ the length of the resonators. With the signal enhancement in the resonators $\beta$ and input power $P_{\rm in}$, the signal to noise ratio is
\begin{equation}
S/N = \frac{P_{\rm in}\beta^2 p(L, \omega) g}{P_{\rm N}},
\end{equation}
where $g$ is a coupling factor of the two resonators and is close to unity in case of symmetrically aligned two Fabry-Perot resonators.
The noise power $P_{\rm N}$ contains integration time $t$ and bandwidth $\Delta\nu$.

\begin{figure} [h]
\begin{center}
  \includegraphics[width=0.8\linewidth]{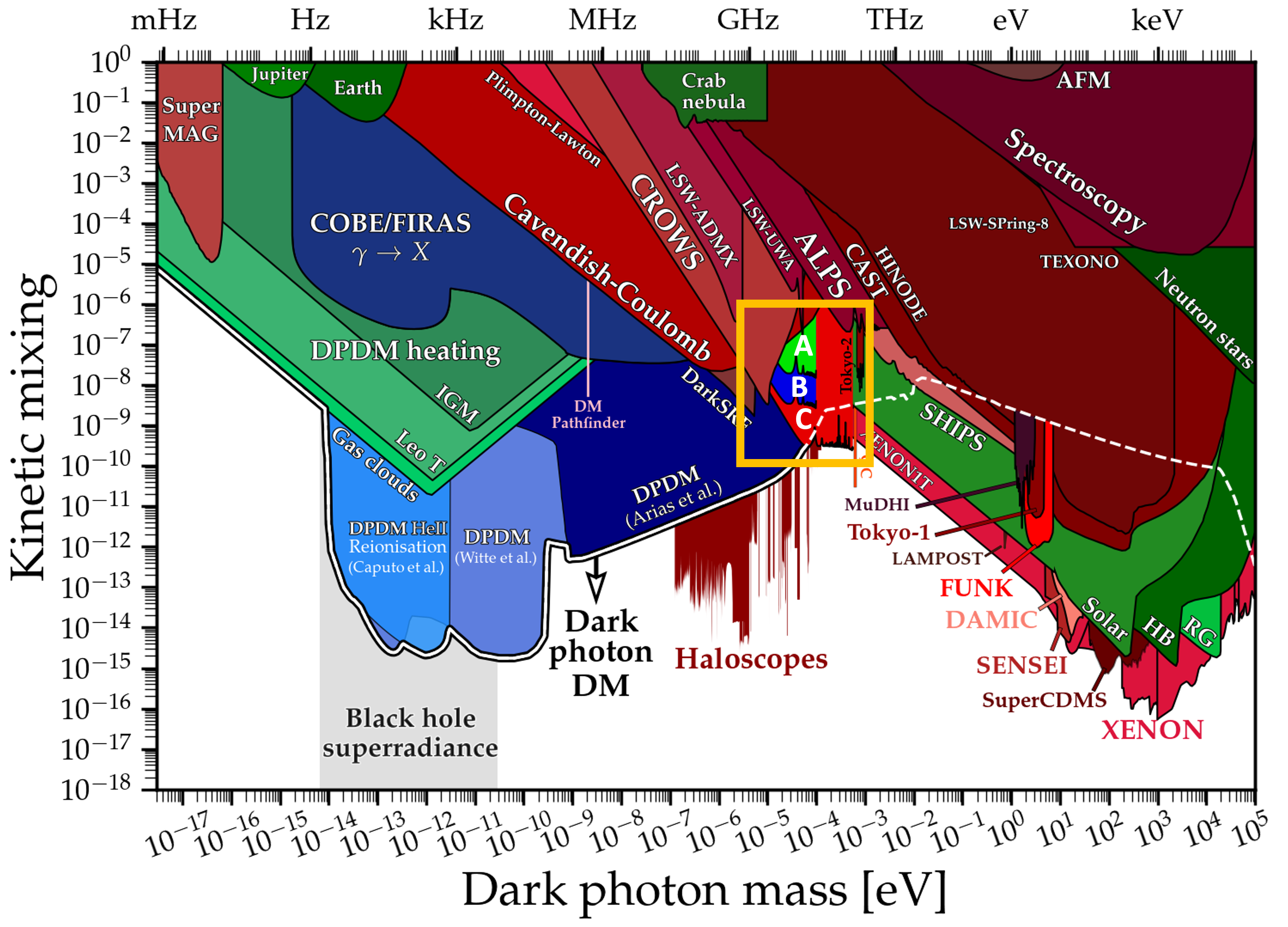}
  \caption{Projection of this work (orange square at the center) on present exclusion limits by using C.~O'Hare's python script~\cite{AxionLimits, PhysRevD.104.095029}. Stage A, B and C show low-power search, high-power search, and single photon search, respectively.}
  \label{fig:limit}
\end{center}
\end{figure}

Figure~\ref{fig:limit} compares the expected exclusion limits ($2\sigma$) of several stages in the proposed dark photon search compared to the existing constraints~\cite{AxionLimits, PhysRevD.104.095029}.
\begin{description}
\item [Stage A: low-power search]
This stage uses the low-power experiment described in Sec.~\ref{sec:PoP_setup}. 
With a commercially available amplifier of $P_{\rm in}=20$~W, two Fabry-Perot resonators with field enhancement factor $\beta=1000$, low-noise amplifier of noise temperature $T_{\rm s}=110$~K, 
we will search the region A (Fig.~\ref{fig:limit}) after data acquisition of $1.5$~hours corresponding to $\Delta\nu_{\rm RBW}=169$~$\mu$Hz. 
Note that three experiments of different resonator lengths ($L=$ $180$, $200$, $220$~mm) are combined to avoid the zeros in Eq.~(\ref{eq:photon-dark-photon-oscillation}).
Thus, the total experimental time will be a few days.
\item [Stage B: high-power search]
This stage is the experiment with the phased-locked gyrotron at KIT described in Sec.~\ref{sec:PLL_gyrotron}.
Instead of using a commercial amplifier, we consider $P_{\rm in}\sim 20$~kW from this 28~GHz gyrotron.
The other conditions are the same as the low-power search.
Cooling down the emitting resonator, which is exposed to 20~kW microwaves, is a challenge.
In a similar Fabry-Perot resonator of 200~GHz, one of the authors handled 1~kW with water cooling~\cite{miyazaki_PhD}.
It is also known that 352~MHz microwaves of 16~kW averaged power can be damped into a relatively compact water-cooled load.
\item [Stage C: single photon at higher frequency]
This stage is the most challenging option with the single photon detector described in Sec.~\ref{sec:JES} combined with the 1~MW gyrotron (Sec.~\ref{sec:PLL_gyrotron}) for ITER, where the ultimate coherency is not necessarily implemented because photon counting is anyway inherently incoherent.
The data acquisition time is three minutes determined by the power supply for the 1~MW-class gyrotron at KIT. 
Since it is not realistic to cool down the emitter resonator exposed to 1~MW power, the Fabry-Perot resonator at the emitter side will be omitted and a travelling wave in a corrugated waveguide will be installed.
The detector-side, including the resonator, must be cooled to below 100~mK for the operation of the single-photon detector.
We also consider the use of a sapphire-based Fabry-Perot resonator that can provide field enhancement of $10^{6}$ in a cryogenic condition for high-frequency microwaves.
\end{description}

\begin{figure} [h]
\begin{center}
  \includegraphics[width=0.8\linewidth]{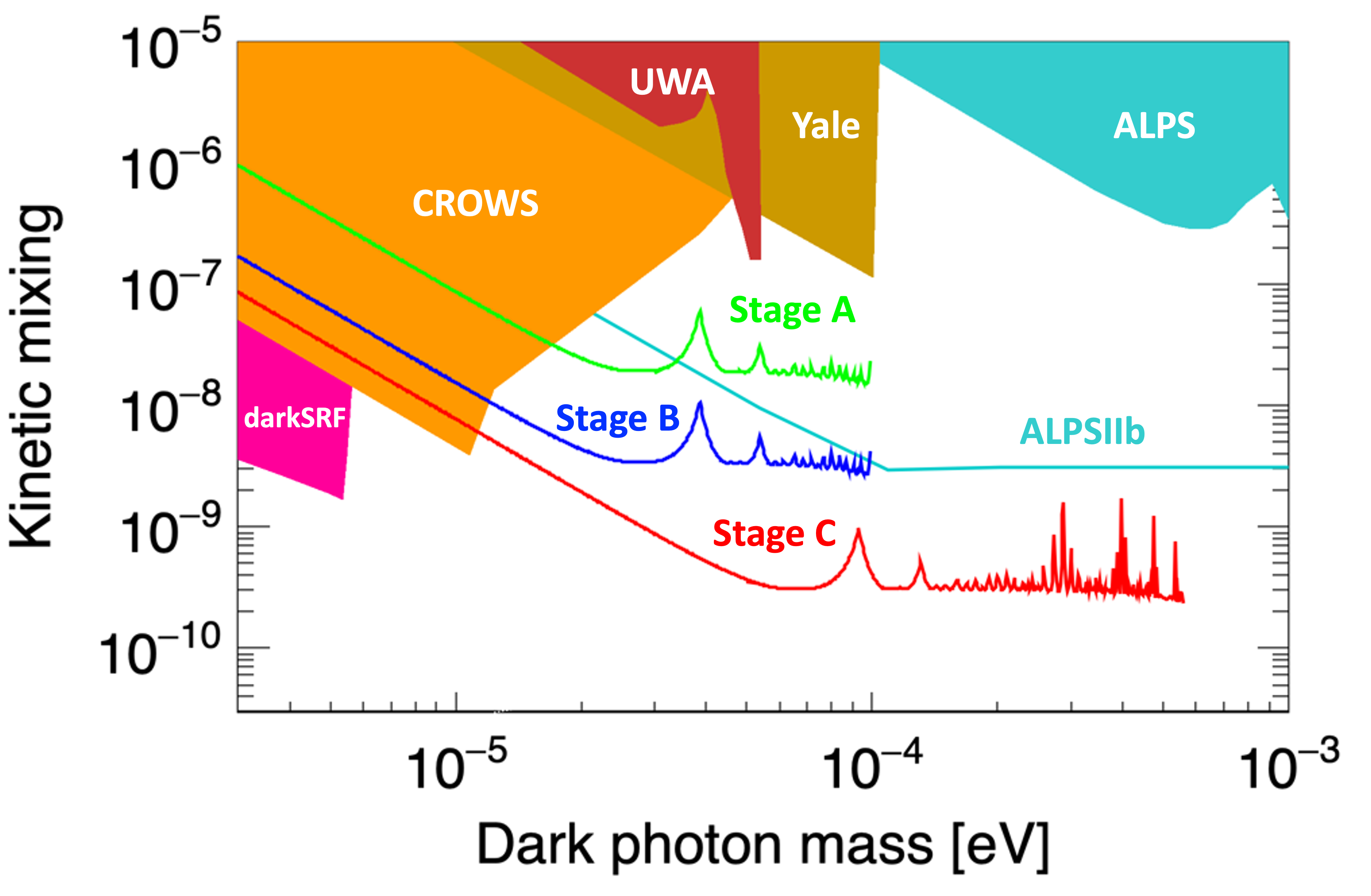}
  \caption{Projection of this work on present LSW exclusion limits and future ALPSII. This work aims at a complementary parameter region to any other LSW experiments.}
  \label{fig:limit_LSW}
\end{center}
\end{figure}

Since this work aims at a new LSW in the parameter range which is complementary to other similar experiments, Figure~\ref{fig:limit_LSW} shows the enlarged view of LSW constraints and our projections.
CROWS~\cite{PhysRevD.88.075014}, UWA~\cite{PhysRevD.87.115008, PhysRevD.88.112004}, and Yale~\cite{SLOCUM201576} are LSW searches with normal conducting RF cavities.
Similar limits with superconducting resonators are indicated as darkSRF~\cite{https://doi.org/10.48550/arxiv.2301.11512}.
ALPS~\cite{EHRET2010149} shows the previous limit of laser LSW while ALPSIIb~~\cite{R_Bahre_2013} is a predicted limit from the upgraded experiment on-going at DESY with infrared lasers and 100~m resonators.
Importantly, even a low-power experiment in Stage A can address the region complementary to the ALPSIIb experiment.
This is because the frequency of this experiment is four orders of magnitude lower than that of ALPSIIb.
Even if the size of the experiment is different, which is 100~m in ALPSIIb while only 20~cm in this proposal,
the use of a different frequency provides a unique opportunity in dark photon physics.
The future haloscope will be operated in the same frequency range as this proposal
but is in principle and in practice complementary to the LSW experiments due to the assumption on dark matter in our galaxy.
Moreover, LSW and haloscope will have technical synergy on the use of millimeter waves.

\subsection{Sub-THz-AXion (STAX) search}
Once the dark photon search is successfully implemented,
axion searches may be performed by adding a dipole magnetic field to the resonators.
The STAX project is a future project of this kind~\cite{CAPPARELLI201637} and this paper gives a prototype of it (proto-STAX).
We develop a single photon sensor in Sec.~\ref{sec:JES} and the 1~MW gyrotron (Sec.~\ref{sec:PLL_gyrotron})
and install the system into superconducting dipole magnets.
We are preliminarily considering the use of two of the Nb$_3$Sn dipoles being developed for the HL-LHC project at CERN.
This magnet will provide an 11~T dipole field inside 60~cm of the magnet bore where the conversion resonators will be placed.
Since the bore is smaller than the diffraction scale of millimeter waves,
we employ Fabry-Perot resonators covered with a corrugated waveguide.
This technique has been developed for the microwave undulator~\cite{PhysRevLett.112.164802} and the application to axion searches is straightforward.
The STAX project is one of the potential successors of the ALPSII experiment.
We need to note that 1~MW power handling will be a significant technical challenge of this project.

\section{Conclusion}
We proposed a new dark photon search in the mass range around $10^{-4}$ to $10^{-3}$~eV by using the LSW technique around 30~GHz.
Wave detection and photon counting are critically compared with respect to the advantage of coherent LSW.
The key is to introduce the quantum coherent state that justifies the semi-classical treatment of photon-dark-photon oscillation.
In a simple setup of a signal generator and a real-time spectrum analyzer,
an excellent relative temporal coherence at 30~GHz was achieved by using the common 10~MHz reference line.
With such a signal coherency, we showed that the thermal noise background can be filtered out with a resolution bandwidth narrower than 1~Hz.
This was achieved by FFT of the waveform over a period longer than 1~second.
We successfully filter that noise power by four orders of magnitude with a narrowest resolution bandwidth of 169~$\mu$Hz.
We revealed a  limitation of noise filtering below this bandwidth probably because of a limited accuracy of the inner circuits of the device. 
With the noise filtering and signal coherency demonstrated in this work, 
we can address the mass region of dark photons that is complementary to other projects by just using commercial amplifiers.
Further improvement is feasible with the recently developed phased-locked gyrotrons that are a promising option toward high-power coherent searches of dark photons.
Although coherency is the huge advantage of this proposal as the initial stage of this project at 30~GHz,
photon counting will surpass the wave detection technique of any coherency in the future
because the coherent wave detection is ultimately limited by the standard quantum limit, and narrow-band filtering showed some limitation around 100~$\mu$Hz.
The development of single photon sensors is on-going by using the nano-wire Josephson junction.
The successful implementation of a dark photon experiment would lead to the future axion project STAX as a successor of infrared axion search of today.


\medskip
\textbf{Acknowledgements} \par 
This work was supported by KIT international excellence fellowship program 2021.
We would like to thank A.~J.~Millar and S.~Albright for useful discussions.

\appendix
\section{Particle counting\label{sec:photon_counting}}
An ideal particle counting of a coherent state provides a probability distribution function $P\left(n; \left| \alpha \right|^2\right)$ from a projection of $| \alpha \rangle$ onto number states $| n\rangle$
\begin{equation}
P\left(n; \left| \alpha \right|^2\right) = \left|\langle n|\alpha \rangle \right|^2 = \frac{\left| \alpha \right|^{2n}}{n!} e^{-\left| \alpha \right|^2},
\end{equation}
which is a Poisson distribution with the mean number of particles at $\left|\alpha\right|^2$.
A conventional experiment in particle physics, such as optical photon counting by a photomultiplier, observes this distribution~\cite{Bellamy:1994bv}.
Practically, photons of frequency higher than optical frequencies, such as X-rays or even $\gamma$-rays, may only be detected as particles.
Detectors are calorimetric and measure photons as energy quanta, event-by-event, typically with some finite energy resolution.
One typical background is dark counts in the sensor or real photons from phenomena other than the signal, such as blackbody radiation.

\section{Wave detection\label{sec:wave_detection}}
The wave detection of a coherent state is well established and is the same as for classical waves even in the sub-quantum regime~\cite{HARTNETT2011346}.
A system composed of an antenna,  linear amplifiers, superheterodyne mixers, and Analog-to-Digital Converter (ADC) directly takes the expectation value of $\hat{E}$ and the consequence is Eq.~(\ref{eq:coherent_expectation}).
Here, the wave amplitude, or more explicitly, voltage and phase information is taken from the quantum system.
If the signal is monochromatic, integration over periods improves the signal-to-noise ratio.
This is a classical description of measuring the expectation value of electric field operators from a coherent state.

The background level of wave detection for microwaves is usually characterized by the noise temperature $T_{\rm s}$.
For simplicity, one models the thermal noise in the Rayleigh–Jeans limit derived from Eq.~(\ref{eq:noise_power_linear}).
This is usually a good approximation because in the microwave range,
\begin{equation}
h\nu < k_{\rm B}T_{\rm s}
\end{equation}
is applied except for the case of standard quantum limit (SQL)~\cite{PhysRevD.88.035020}
\begin{equation}
k_{\rm B}T_{\rm SQL} = h\nu,
\end{equation}
which is from the quantum uncertainty (Robertson-Schr\"{o}dinger inequality) of the signal itself.
For 30~GHz microwaves, we have $T_{\rm SQL} = 1.4$~K but practically the measurement is blocked by $T_{\rm s}\sim 7$~K from a commercially available low-noise amplifier.
Going down to $T_{\rm SQL}$ was realized and has been used for dark matter search below 15~GHz with a Josephson parametric amplifier (JPA)~\cite{2013APS..APRQ11007A}.
Parametric amplifiers above 20~GHz are under development.
Since $T_{\rm SQL}$ is relatively high for 30~GHz, one does not gain much signal-to-noise ratio by cooling the system below 1~K with a dilution refrigerator.

\section{Coherency of dark photons search}\label{sec:dark_photon_coherency}
Ref.~\cite{PhysRevD.88.035020} investigated the use of classical-wave detection and photon counting techniques in the context of dark matter haloscope experiments, in which dark photons are converted to microwaves inside a resonant cavity with a loaded quality factor $Q_{\rm c}$.
In the cold dark matter model, the velocities of dark photons in our galaxy have a Maxwell-Boltzmann distribution.
This makes a finite bandwidth $\Delta\nu_{\rm \gamma'}$ in the microwave signal $\nu$ characterized by a signal quality factor $Q_{\it \gamma'}=\nu/\Delta\nu_{\rm \gamma'}\sim10^{6}$.
We follow the discussion in Ref.~\cite{PhysRevD.88.035020}, in which fake signals, such as dark counts or cross-talks, are disregarded for simplicity.
The noise power of classical wave detection can be given by the Johnson-Nyquist noise which satisfies~\cite{PhysRev.32.110, PhysRev.83.34}
\begin{equation}\label{eq:noise_power_linear}
P_{\rm l} = h\nu \left(\bar{n} + 1 \right) \sqrt{\frac{\Delta\nu_{\gamma'}}{t}},
\end{equation}
where $t$ is the data taking time and $\bar{n}$ is the Planck distribution
\begin{equation}
\bar{n} = \frac{1}{e^{h\nu/k_{\rm B}T}-1}
\end{equation}
of system temperature $T$.
Note that $\bar{n}+1$ represents the quantum effect in the Johnson-Nyquist noise and this gives the standard quantum limit of wave detection.
The noise power of photon sensors is given by counting statistics
\begin{equation}\label{eq:noise_power_single}
P_{\rm sp} = h\nu \sqrt{\frac{\eta \bar{n}Q_{\rm c}}{2\pi \nu t}},
\end{equation}
where $\eta$ is quantum efficiency of the photon sensor.
The ratio of $P_{\rm l}$ and $P_{\rm sp}$ can be written as
\begin{equation}
\frac{P_{\rm l}}{P_{\rm sp}} = \frac{1}{\sqrt{2\pi\eta}} \left(\sqrt{\bar{n}} + \frac{1}{\sqrt{\bar{n}}} \right) \sqrt{\frac{Q_{\rm c}}{Q_{\gamma'}}}. \label{eq:ratio_noise_power}
\end{equation}
The authors~\cite{PhysRevD.88.035020} found that $P_{\rm sp}$ is lower than $P_{\rm l}$ at dark photon masses above 41~$\mu$eV if the photo-sensor is cooled down below 100~mK.
The recent progress in developing single photon detectors may soon fulfill these requirements~\cite{PhysRevApplied.14.034055}.

This conclusion needs to be reconsidered for dark photons generated in LSW.
Unlike dark matter searches, one can enhance the temporal coherency or even lock the phase of the signal, and consequently, the signal $Q_{\gamma'}$ can be much higher than that of dark matter ($10^6$).
In this case, Eq.~(\ref{eq:noise_power_linear}) leads to very low $P_{\rm l}$.
Therefore, the cross-over point in Eq.~(\ref{eq:ratio_noise_power}) can be at even higher frequency ($\gg 10$~GHz) and even lower temperature ($\ll 100$~mK) than the case of dark matter dark photons.
Implementation of single photon sensors is not justified for LSW until one reseaches this technically challenging cross-over point.

\medskip

%
\bibliographystyle{MSP}
\bibliography{dark_photon}

\begin{thebibliography}{10}
\providecommand{\url}[1]{\texttt{#1}}
\providecommand{\urlprefix}{URL }

\bibitem{201230}
{{C}{M}{S} collaboration},
\newblock \emph{Physics Letters B} \textbf{2012}, \emph{716}, 1 30.

\bibitem{20121}
{{A}{T}{L}{A}{S} collaboration},
\newblock \emph{Physics Letters B} \textbf{2012}, \emph{716}, 1 1.

\bibitem{PhysRevD.79.015014}
N.~Arkani-Hamed, D.~P. Finkbeiner, T.~R. Slatyer, N.~Weiner,
\newblock \emph{Phys. Rev. D} \textbf{2009}, \emph{79} 015014.

\bibitem{Nima-Arkani-Hamed_2008}
N.~Arkani-Hamed, N.~Weiner,
\newblock \emph{Journal of High Energy Physics} \textbf{2008}, \emph{2008}, 12
  104.

\bibitem{PhysRevLett.101.231301}
J.~L. Feng, J.~Kumar,
\newblock \emph{Phys. Rev. Lett.} \textbf{2008}, \emph{101} 231301.

\bibitem{HOLDOM1986196}
B.~Holdom,
\newblock \emph{Physics Letters B} \textbf{1986}, \emph{166}, 2 196.

\bibitem{1970ApJ...159..379R}
V.~C. {Rubin}, J.~{Ford}, W.~Kent,
\newblock \emph{Astrophysical Journal} \textbf{1970}, \emph{159} 379.

\bibitem{Markevitch_2004}
M.~Markevitch, A.~H. Gonzalez, D.~Clowe, A.~Vikhlinin, W.~Forman, C.~Jones,
  S.~Murray, W.~Tucker,
\newblock \emph{The Astrophysical Journal} \textbf{2004}, \emph{606}, 2 819.

\bibitem{Clowe_2004}
D.~Clowe, A.~Gonzalez, M.~Markevitch,
\newblock \emph{The Astrophysical Journal} \textbf{2004}, \emph{604}, 2 596.

\bibitem{6380414}
T.~Suehara, K.~Owada, A.~Miyazaki, T.~Yamazaki, S.~Asai, T.~Kobayashi,
\newblock In \emph{2012 37th International Conference on Infrared, Millimeter,
  and Terahertz Waves}. \textbf{2012} 1--2.

\bibitem{9370386}
A.~Miyazaki, P.~Spagnolo,
\newblock In \emph{2020 45th International Conference on Infrared, Millimeter,
  and Terahertz Waves (IRMMW-THz)}. \textbf{2020} 1--2.

\bibitem{Isleif2022}
K.-S. Isleif, A.~Collaboration),
\newblock \emph{Moscow University Physics Bulletin} \textbf{2022}, \emph{77}
  120.

\bibitem{PhysRevD.99.022001}
Z.~R. Bush, S.~Barke, H.~Hollis, A.~D. Spector, A.~Hallal, G.~Messineo, D.~B.
  Tanner, G.~Mueller,
\newblock \emph{Phys. Rev. D} \textbf{2019}, \emph{99} 022001.

\bibitem{https://doi.org/10.48550/arxiv.2010.02334}
A.~Hallal, G.~Messineo, M.~D. Ortiz, J.~Gleason, H.~Hollis, D.~B. Tanner,
  G.~Mueller, A.~Spector,
\newblock The heterodyne sensing system for the alps ii search for sub-ev
  weakly interacting particles, \textbf{2020},
\newblock \urlprefix\url{https://arxiv.org/abs/2010.02334}.

\bibitem{Caspers:1195741}
F.~Caspers, S.~Federmann, D.~Seebacher \textbf{2009}.

\bibitem{F_Caspers_2009}
F.~Caspers, J.~Jaeckel, A.~Ringwald,
\newblock \emph{Journal of Instrumentation} \textbf{2009}, \emph{4}, 11 P11013.

\bibitem{PhysRevD.88.075014}
M.~Betz, F.~Caspers, M.~Gasior, M.~Thumm, S.~W. Rieger,
\newblock \emph{Phys. Rev. D} \textbf{2013}, \emph{88} 075014.

\bibitem{JAECKEL2008509}
J.~Jaeckel, A.~Ringwald,
\newblock \emph{Physics Letters B} \textbf{2008}, \emph{659}, 3 509.

\bibitem{PhysRevLett.61.2285}
D.~F. Bartlett, S.~L\"ogl,
\newblock \emph{Phys. Rev. Lett.} \textbf{1988}, \emph{61} 2285.

\bibitem{1955PhRv...97.1387G}
M.~{Gell-Mann}, A.~{Pais},
\newblock \emph{Physical Review} \textbf{1955}, \emph{97}, 5 1387.

\bibitem{1968JETP...26..984P}
B.~{Pontecorvo},
\newblock \emph{Soviet Journal of Experimental and Theoretical Physics}
  \textbf{1968}, \emph{26} 984.

\bibitem{Schrodinger1926}
E.~{Schr\"{o}dinger},
\newblock \emph{Naturwissenschaften} \textbf{1926}, \emph{14} 664.

\bibitem{PhysRev.131.2766}
R.~J. Glauber,
\newblock \emph{Phys. Rev.} \textbf{1963}, \emph{131} 2766.

\bibitem{9567608}
A.~Miyazaki,
\newblock In \emph{2021 46th International Conference on Infrared, Millimeter
  and Terahertz Waves (IRMMW-THz)}. \textbf{2021} 1--2.

\bibitem{Miyazaki:2798112}
A.~Miyazaki, F.~Caspers, J.~L. Steinmann, T.~Ruess, J.~Jelonnek \textbf{2021}.

\bibitem{9896105}
E.~Br\"{u}ndermann, J.~L. Steinmann, I.~Morohashi, S.~Nakajima, S.~Saito,
  N.~Sekine, A.-S. M\"{u}ller, I.~Hosako,
\newblock In \emph{2022 47th International Conference on Infrared, Millimeter
  and Terahertz Waves (IRMMW-THz)}. \textbf{2022} 1--2.

\bibitem{KARA}
Karlsruhe research accelerator ({KARA}),
\newblock \url{https://www.ibpt.kit.edu/kara.php},
\newblock Accessed: 2022-11-09.

\bibitem{https://doi.org/10.1049/iet-map.2018.0103}
L.~Wang, K.~Dong, W.~He, J.~Wang, Y.~Luo, A.~W. Cross, K.~Ronald, A.~D.~R.
  Phelps,
\newblock \emph{IET Microwaves, Antennas \& Propagation} \textbf{2018},
  \emph{12}, 11 1752.

\bibitem{doi:10.1063/1.5144590}
L.~J.~R. Nix, L.~Zhang, W.~He, C.~R. Donaldson, K.~Ronald, A.~W. Cross, C.~G.
  Whyte,
\newblock \emph{Physics of Plasmas} \textbf{2020}, \emph{27}, 5 053101.

\bibitem{Thumm2020}
M.~Thumm,
\newblock \emph{J Infrared Milli Terahz Waves} \textbf{2020}, \emph{41} 1.

\bibitem{Malygin2014}
A.~Malygin, S.~Illy, I.~Pagonakis, K.~Avramidis, M.~Thumm, J.~Jelonnek,
  L.~Ives, G.~Collins,
\newblock In \emph{Proceedings of the 9th International Workshop ``Strong
  Microwaves and Terahertz Waves: Sources and Applications''}. \textbf{2014}
  162--163.

\bibitem{miyazaki_PhD}
A.~Miyazaki,
\newblock Ph.D. thesis, The University of Tokyo (Springer Thesis),
  \textbf{2014}.

\bibitem{Denisov2022}
G.~Denisov, A.~Kuftin, A.~Chirkov, M.~Bakulin, E.~Soluyanova, E.~Tai,
  G.~Gobobyatnikov, R.~Morozkin, B.~Moyshevich, M.~Glyavin,
\newblock In \emph{Proceedings of the 23rd IEEE International Vacuum
  Electronics Conference (IVEC 2022)}. \textbf{2022} 2--4.

\bibitem{PhysRevApplied.14.034055}
F.~Paolucci, N.~Ligato, V.~Buccheri, G.~Germanese, P.~Virtanen, F.~Giazotto,
\newblock \emph{Phys. Rev. Applied} \textbf{2020}, \emph{14} 034055.

\bibitem{doi:10.1063/5.0021996}
F.~Paolucci, V.~Buccheri, G.~Germanese, N.~Ligato, R.~Paoletti, G.~Signorelli,
  M.~Bitossi, P.~Spagnolo, P.~Falferi, M.~Rajteri, C.~Gatti, F.~Giazotto,
\newblock \emph{Journal of Applied Physics} \textbf{2020}, \emph{128}, 19
  194502.

\bibitem{AxionLimits}
C.~O'Hare,
\newblock cajohare/axionlimits: Axionlimits,
\newblock \url{https://cajohare.github.io/AxionLimits/}, \textbf{2020}.

\bibitem{PhysRevD.104.095029}
A.~Caputo, A.~J. Millar, C.~A.~J. O'Hare, E.~Vitagliano,
\newblock \emph{Phys. Rev. D} \textbf{2021}, \emph{104} 095029.

\bibitem{PhysRevD.87.115008}
S.~R. Parker, G.~Rybka, M.~E. Tobar,
\newblock \emph{Phys. Rev. D} \textbf{2013}, \emph{87} 115008.

\bibitem{PhysRevD.88.112004}
S.~R. Parker, J.~G. Hartnett, R.~G. Povey, M.~E. Tobar,
\newblock \emph{Phys. Rev. D} \textbf{2013}, \emph{88} 112004.

\bibitem{SLOCUM201576}
P.~Slocum, O.~Baker, J.~Hirshfield, Y.~Jiang, A.~Malagon, A.~Martin,
  S.~Shchelkunov, A.~Szymkowiak,
\newblock \emph{Nuclear Instruments and Methods in Physics Research Section A:
  Accelerators, Spectrometers, Detectors and Associated Equipment}
  \textbf{2015}, \emph{770} 76.

\bibitem{https://doi.org/10.48550/arxiv.2301.11512}
A.~Romanenko, R.~Harnik, A.~Grassellino, R.~Pilipenko, Y.~Pischalnikov, Z.~Liu,
  O.~S. Melnychuk, B.~Giaccone, O.~Pronitchev, T.~Khabiboulline, D.~Frolov,
  S.~Posen, A.~Berlin, A.~Hook,
\newblock New exclusion limit for dark photons from an srf cavity-based search
  (dark srf), \textbf{2023},
\newblock \urlprefix\url{https://arxiv.org/abs/2301.11512}.

\bibitem{EHRET2010149}
K.~Ehret, M.~Frede, S.~Ghazaryan, M.~Hildebrandt, E.-A. Knabbe, D.~Kracht,
  A.~Lindner, J.~List, T.~Meier, N.~Meyer, D.~Notz, J.~Redondo, A.~Ringwald,
  G.~Wiedemann, B.~Willke,
\newblock \emph{Physics Letters B} \textbf{2010}, \emph{689}, 4 149.

\bibitem{R_Bahre_2013}
R.~Bähre, B.~D\"{o}brich, J.~Dreyling-Eschweiler, S.~Ghazaryan, R.~Hodajerdi,
  D.~Horns, F.~Januschek, E.~A. Knabbe, A.~Lindner, D.~Notz, A.~Ringwald, J.~E.
  von Seggern, R.~Stromhagen, D.~Trines, B.~Willke,
\newblock \emph{Journal of Instrumentation} \textbf{2013}, \emph{8}, 09 T09001.

\bibitem{CAPPARELLI201637}
L.~Capparelli, G.~Cavoto, J.~Ferretti, F.~Giazotto, A.~Polosa, P.~Spagnolo,
\newblock \emph{Physics of the Dark Universe} \textbf{2016}, \emph{12} 37.

\bibitem{PhysRevLett.112.164802}
S.~Tantawi, M.~Shumail, J.~Neilson, G.~Bowden, C.~Chang, E.~Hemsing,
  M.~Dunning,
\newblock \emph{Phys. Rev. Lett.} \textbf{2014}, \emph{112} 164802.

\bibitem{Bellamy:1994bv}
E.~H. Bellamy, G.~Bellettini, F.~Gervelli, M.~Incagli, D.~Lucchesi,
  C.~Pagliarone, F.~Zetti, Y.~Budagov, I.~Chirikov-Zorin, S.~Tokar,
\newblock \emph{Nucl. Instrum. Meth. A} \textbf{1994}, \emph{339} 468.

\bibitem{HARTNETT2011346}
J.~G. Hartnett, J.~Jaeckel, R.~G. Povey, M.~E. Tobar,
\newblock \emph{Physics Letters B} \textbf{2011}, \emph{698}, 5 346.

\bibitem{PhysRevD.88.035020}
S.~K. Lamoreaux, K.~A. van Bibber, K.~W. Lehnert, G.~Carosi,
\newblock \emph{Phys. Rev. D} \textbf{2013}, \emph{88} 035020.

\bibitem{2013APS..APRQ11007A}
M.~A. {Anil},
\newblock In \emph{APS April Meeting Abstracts}, volume 2013 of \emph{APS
  Meeting Abstracts}. \textbf{2013} Q11.007.

\bibitem{PhysRev.32.110}
H.~Nyquist,
\newblock \emph{Phys. Rev.} \textbf{1928}, \emph{32} 110.

\bibitem{PhysRev.83.34}
H.~B. Callen, T.~A. Welton,
\newblock \emph{Phys. Rev.} \textbf{1951}, \emph{83} 34.

\end{thebibliography}



\end{document}